\documentclass[twocolumn,preprintnumbers,amsmath,amssymb]{revtex4}
\usepackage{graphicx}% Include figure files
\usepackage{dcolumn}% Align table columns on decimal point
\usepackage{bm}% bold math

\begin{document}
\title{Self-passivation of vacancies in $\alpha$-PbO.}
\author{J. Berashevich, J.A. Rowlands}
\affiliation{Thunder Bay Regional Research Institute, 290 Munro St., Thunder Bay, ON, P7B 5E1, Canada}
\author{A. Reznik}
\affiliation{Thunder Bay Regional Research Institute, 290 Munro St., Thunder Bay, ON, P7B 5E1, Canada}
\affiliation{Department of Physics, Lakehead University, 955 Oliver Road, Thunder Bay, ON, P7B 5E1}

\begin{abstract}
We introduce a self-passivation of single lead (Pb) and oxygen (O) vacancies
in the $\alpha$-PbO compound through formation of a Pb-O vacancy pair.
The preferential mechanism for pair formation involves initial development
of the single Pb vacancy which, by weakening the covalent bonding,
sets up the crystal lattice for an appearance of the O vacancy.
Binding of the Pb and O vacancies occurs through the ionization interactions.
Since no dangling bonds appear at the Pb-O pair site,
this defect has a minor effect on the electronic properties.
In such, vacancy self-passivation offers a practical way
to improve the transport properties in thermally grown PbO layers.
\end{abstract}

\maketitle
Polycrystalline Lead Oxide (PbO) is one of the few photoconductive
materials with a long- more than 60 years - history of employment
in imaging devices. Although PbO is extensively used as a photoconductor,
very little is known about the electronic properties and charge transport in this material.
It is generally believed that transport in PbO is controlled by
trapping in defects and that trapping is a cause of low mobility-time product \cite{rowland}.
However, the nature of defects is not fully understood. Emerging
applications of PbO in the direct conversion flat panel radiation
medical imaging detectors revived the interest in studying defects
in this material as defects can significantly affect imaging performance \cite{simon}.

Our recent modeling of the native point defects in $\alpha$-PbO \cite{defects} has shown 
that thermally deposited PbO layers should contain a significant
concentration of single vacancies due to their moderate formation energies.
Single vacancies are amphoteric defects appearing in the different charge states. 
In the neutral charge state, the O vacancy ($V^{\operatorname{O}(0)})$) 
holds two electrons and forms the deep donor level. 
The neutral Pb vacancy ($V^{\operatorname{Pb}(0)})$) is filled with holes, and is a shallow acceptor. 
It was established that these vacancies prefer to appear doubly ionized, 
$V^{\operatorname{Pb}(2-)}$ and $V^{\operatorname{O}(2+)}$, acting as compensating centres to each other. 
Indeed, two compensating vacancies have a lower formation energy 
than neutral ones, such as $\Delta E^f(V^{\operatorname{Pb}(0)}+V^{\operatorname{O}(0)})$-
$\Delta E^f(V^{\operatorname{Pb}(2-)}+V^{\operatorname{O}(2+)})$=0.78 eV \cite{defects}. 

The fact that vacancies prefer to appear in charge states suggests
that we have to consider the ionization interactions between them and
the formation of a neutral vacancy pair $V^{\operatorname{Pb-O}}$
instead of two separate compensating vacancies.
This should further decrease the formation energy,
approximately speaking, the free energy required to
insert a defect in a lattice is reduced by the energy liberated due
to ionization of the donor and acceptor. Indeed, the formation of the defect
complexes in favor over the single defects is 
often observed during material deposition \cite{dutta}. In this work, 
we present our study of the formation mechanism of $V^{\operatorname{Pb-O}}$
pair in PbO layers and its effect on the electronic properties. 

\begin{figure}
\includegraphics[scale=0.23]{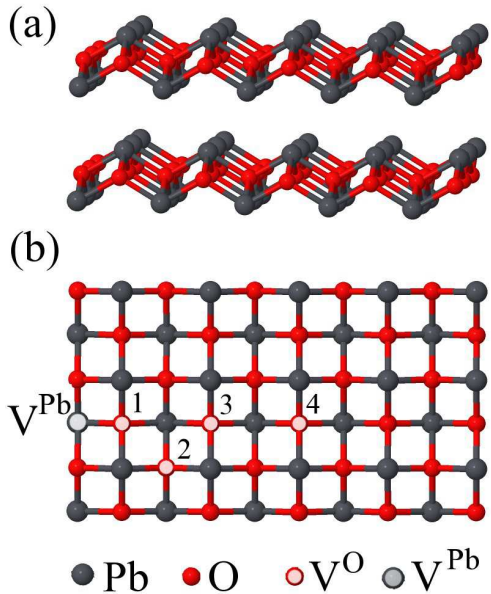}
\caption{\label{fig:fig1} Colour on-line. (a) The crystal structure of $\alpha$-PbO is
shown for supercell of size 5$\times$3$\times$2. 
(b) Scheme demonstrates a position of the O vacancy 
(numbers 1,2,3 and 4) relative to the Pb vacancy to generate the 
$V^{\operatorname{Pb-O}}$ vacancy pair.}
\end{figure}

Analysis of the formation of $V^{\operatorname{Pb-O}}$ vacancy
pair was performed using the density functional theory (DFT)
available in the Wien2k package \cite{wien}
which utilizes the full-potential augmented plane-wave method.
The Perdew-Burke-Ernzerhof parameterization \cite{GGA} of the generalized gradient approximation (GGA)
to DFT has been implemented.
We assigned only $5p$, $5d$, $6s$ and $6p$ electrons of the Pb atom and $2s$ and $2p$
electrons of the O atom to the valence states,
while the lower energy electrons were included into the core shells.
The supercell approach was applied in which the supercell was taken of 120 atom size 
(5$\times$3$\times$2 array of the primitive cells shown in Fig.~\ref{fig:fig1} (a)). 
For some calculations, the supercell of 108 atom size has been used (3$\times$3$\times$3).
The supercell has been optimized with respect to the internal
degrees of freedom, for which minimization based on the forces has been applied \cite{forces}. 
The following parameters have been used: $RK_{max}$=7 (product of the atomic sphere
radius and the plane-wave cut-off in k-space),
and standard Monkhorst-Pack mesh of size 4$\times$4$\times$2 in k-space for Brillouin zone.
Assuming that PbO layers are composed primarily of $\alpha$-PbO single crystals \cite{book},
we have considered it as a model compound. The $\alpha$-PbO single crystal is
of the tetragonal symmetry (129P4/nmm), its crystal structure is 
layered and the layers are held together due to interlayer orbital
overlap of the Pb:$6s^2$ lone pairs \cite{terp}. The detailed information on
structure and parameters used in our investigation of the vacancy defects can be found elsewhere \cite{defects}.

Upon formation of the vacancy pair, deficiency of two electrons at the Pb vacancy
is compensated by two electrons occupying the O vacancy and
dangling bonds are closed on each other.
Since Pb vacancy state is delocalized beyond the vacancy site
(its wavefunction is spread over at least five nearest-neighbours),
it is not required for O and Pb vacancies to be located at the nearest sites to form a pair:
schematic presentation of a position of the interacting vacancies relative to each other is shown in Fig.~\ref{fig:fig1} (b). 
The electronic interactions between vacancies are
calculated as function of separation distance.
Our analysis has revealed that an extra electron or hole added to the pair is delocalized implying that $V^{\operatorname{Pb-O}}$ 
has only zero charge state for which the formation energy is \cite{walle}:
\begin{equation}
\Delta E^f(V^{\operatorname{Pb-O}})=E_{tot}(D)-E_{tot}(X)+\mu_{\operatorname{(O)}}+\mu_{\operatorname{(Pb)}}
\label{eq:one}
\end{equation}
where $E_{tot}(D)$ and $E_{tot}(X)$ are the total energy of
the system containing the defect and defect-free, respectively;
$\mu_{\operatorname{(O)}}=E_{tot}(\operatorname{O})+\mu_{\operatorname{(O)}}^*$ and 
$\mu_{\operatorname{(Pb)}}=E_{tot}(\operatorname{Pb})+\mu_{\operatorname{(Pb)}}^*$,
are the chemical potentials of removed atoms (O and Pb atoms) where $\mu_i^*$ is
the energy required to move atom from vacuum to the conditions of growth
($\mu_{\operatorname{(Pb)}}^*+\mu_{\operatorname{(O)}}^*=\Delta_f H^0(\operatorname{PbO})=$-2.92 eV \cite{defects}
is defined for the stable compound).

\begin{figure}
\includegraphics[scale=0.34]{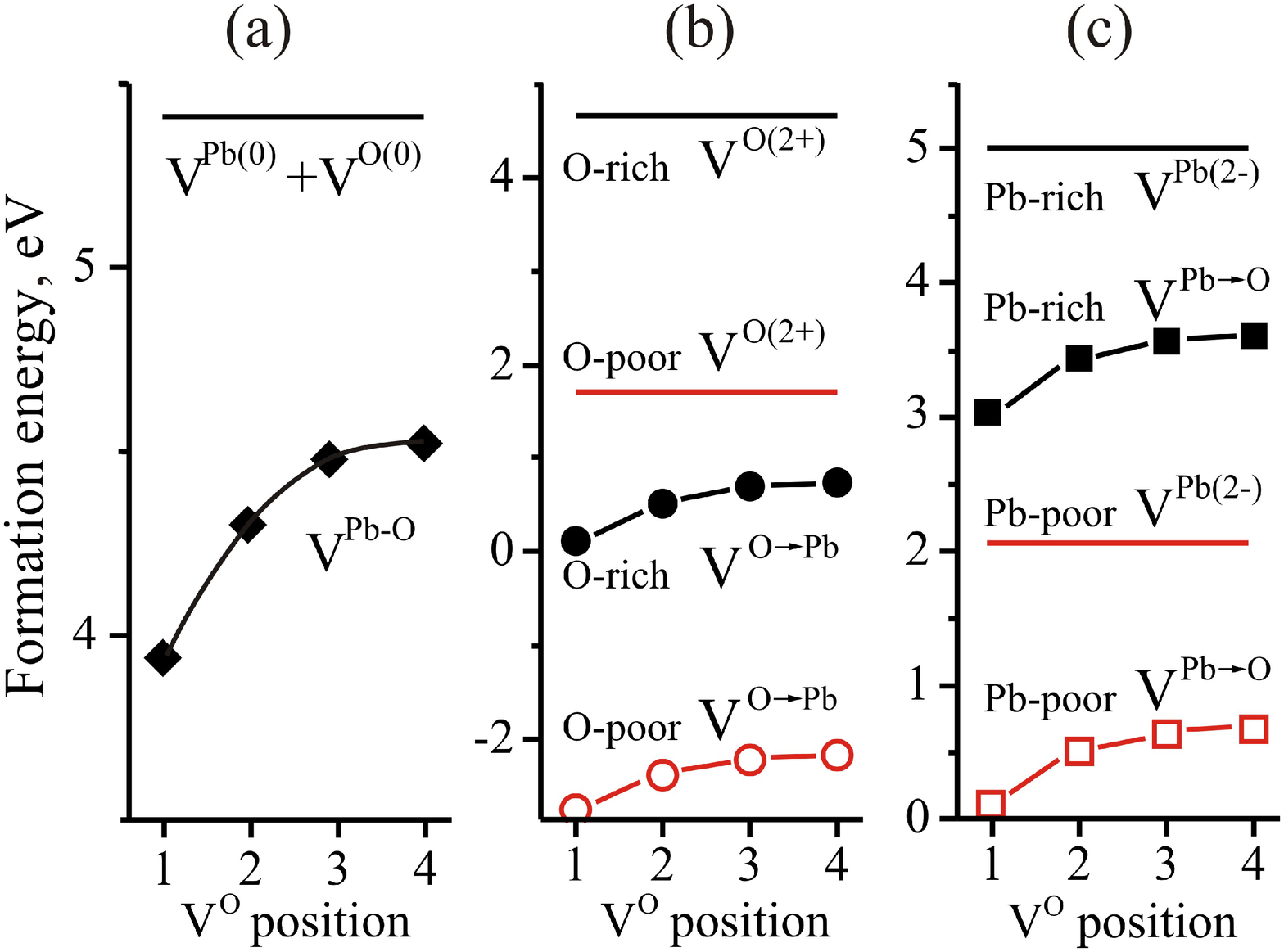}
\caption{\label{fig:fig2} The formation energy of the vacancy pair
as a function of the vacancy separation 
($V^{\operatorname{O}}$ position is referred to positions 1,2,3 and 4 in Fig.~\ref{fig:fig1} (b)):
(a) formation of the $V^{\operatorname{Pb-O}}$ vacancy pair. 
(b) binding of $V^{\operatorname{O}}$ to previously formed
$V^{\operatorname{Pb}}$ ($V^{\operatorname{O\rightarrow Pb}}$).
(c) binding of $V^{\operatorname{Pb}}$ to 
already existing $V^{\operatorname{O}}$ ($V^{\operatorname{Pb\rightarrow O}}$). 
The formation energy of the single charged vacancies are calculated elsewhere \cite{defects}.}
\end{figure}

To account for ionization interactions between vacancies
we have analyzed the non-interacting and interacting vacancies.
For non-interacting case, $V^{\operatorname{O}}$ and $V^{\operatorname{Pb}}$ 
have been placed into the same supercell, but in different layers.
The interactions between such vacancies are absent and each
vacancy creates its own localized state, $V^{\operatorname{O}(0)}$ and $V^{\operatorname{Pb}(0)}$. 
The formation energy under typical growth conditions 
($\mu_{\operatorname{(Pb)}}^*+\mu_{\operatorname{(O)}}^*=$-2.92 eV)
is found to be $\Delta E^f(V^{\operatorname{O}(0)}+V^{\operatorname{Pb}(0)})$=5.41 eV.
To initiate the interaction between vacancies,
they have been placed in the same supercell and the same layer.
The formation energies of the vacancy pair $\Delta E^f(V^{\operatorname{Pb-O}})$
as function of the vacancy separation are presented in Fig.~\ref{fig:fig2} (a) 
(in the following we assume that $\Delta E^f(V^{\operatorname{Pb-O}})$ is saturated at position 4).

The total lowering in the formation energy upon vacancy binding is found to be 
$E_{bind}$=-1.47 eV (vacancies are located next to each other) 
which suggests that $V^{\operatorname{Pb-O}}$ is extremely energetically favored. 
There are two primary parts contributing to $E_{bind}$: interaction energy $E_{int}$ and lattice relaxation $E_{lat}$. 
The $E_{int}$ is the energy change induced by the electron exchange between
$V^{\operatorname{O}(0)}$ and $V^{\operatorname{Pb}(0)}$ causing their ionization to 
$V^{\operatorname{Pb}(2-)}$ and $V^{\operatorname{O}(2+)}$ states which is followed by lattice relaxation $E_{lat}$. 
The value of $E_{int}$ can be determined through the difference 
between the considered position and the point of saturation (position 4 is considered here to be the saturation point), 
is equal to 0.55 eV for $V^{\operatorname{O}}$ in position 1. 
$E_{int}$ gradually decreases as oxygen vacancy is moved away (see Fig.~\ref{fig:fig2} (a)) 
thus increasing the formation energy of a pair. 
The magnitude of $E_{int}$=0.55 eV is foreseen to be underestimated
because the state formed by the $V^{\operatorname{Pb}}$ vacancy is largely
delocalized and when $V^{\operatorname{O}}$ is placed at position 4, 
it already may interact with the periodic copies of $V^{\operatorname{Pb}}$. 
To eliminate this effect, the supercell of larger size has to 
be implemented which is not computationally affordable for us. 
For all positions depicted in Fig.~\ref{fig:fig1} (b), 
$E_{lat}$ is constant and equals to 0.92 eV as the charge state within $V^{\operatorname{Pb}}$ is not changed.

Although vacancy binding lowers the formation energy, 
its value is still high ($\geq$ 4 eV) because the Pb and O vacancies are considered to appear simultaneously. 
The more realistic case appears to be settling as the compensating
vacancy originates near already existing vacancy. 
We introduce notation $V^{\operatorname{Pb\rightarrow O}}$ when $V^{\operatorname{Pb}}$ 
is formed nearby existing $V^{\operatorname{O}}$, and $V^{\operatorname{O\rightarrow Pb}}$ for the opposite case. 
The formation energies are:
\begin{equation}
\Delta E^f(V^{\operatorname{O\rightarrow Pb}})=E_{tot}(D)-E_{tot}(\operatorname{Pb})+\mu_{\operatorname{(O)}}
\label{eq:two}
\end{equation}
\begin{equation}
\Delta E^f(V^{\operatorname{Pb\rightarrow O}})=E_{tot}(D)-E_{tot}(\operatorname{O})+\mu_{\operatorname{(Pb)}}
\label{eq:three}
\end{equation}
where $E_{tot}(\operatorname{Pb})$ and $E_{tot}(\operatorname{O})$ are the total energy of the 
system containing the single vacancy $V^{\operatorname{Pb}}$ or $V^{\operatorname{O}}$, respectively. 

We have plotted the results on the formation energy of the vacancy pairs,
$V^{\operatorname{O\rightarrow Pb}}$
and $V^{\operatorname{Pb\rightarrow O}}$, in Fig.~\ref{fig:fig2} (b) and (c), respectively.
The different growth conditions have been considered:
$\mu_{\operatorname{(Pb)}}^*$=0.0 eV and $\mu_{\operatorname{(O)}}^*$=-2.92 eV
are defined for Pb-rich/O-poor limit,
while $\mu_{\operatorname{(Pb)}}^*$=-2.92 eV and $\mu_{\operatorname{(O)}}^*$=0.0 eV
for Pb-poor/O-rich limit \cite{defects}.
There is a very important trend to be observed in Fig.~\ref{fig:fig2} (b) and (c):
the values of $\Delta E^f(V^{\operatorname{Pb\rightarrow O}})$ and
$\Delta E^f(V^{\operatorname{O\rightarrow Pb}})$ are significantly lower than
the formation energy of the corresponding single vacancies in their doubly ionized state \cite{defects}.
Taking into account that binding energy is only $E_{bind}$ =-1.47 eV,
this trend indicates that the lattice distortion induced by the single
vacancy reduces the energy required to generate another vacancy nearby.
The single vacancy disturbs the lattice periodicity that alters the
atomic forces between atoms and weakens the covalent bonds around the vacancy site.

Therefore, the Pb vacancy sets up a lattice to accommodate the O vacancy, 
thus causing a reduction in the formation energy
$\Delta E^f(V^{\operatorname{O}(2+)})-\Delta E^f(V^{\operatorname{O\rightarrow Pb}})$=4.52 eV 
($V^{\operatorname{O}}$ is in position 1). 
Negative $\Delta E^f(V^{\operatorname{O\rightarrow Pb}})$ means that the lattice containing
$V^{\operatorname{Pb}}$ tends to "suck in" O vacancy and formation of 
$V^{\operatorname{O\rightarrow Pb}}$ pair becomes spontaneous. 
We expect the Pb vacancies to be completely passivated by O vacancies, 
as even for O-rich limit the formation energy $\Delta E^f(V^{\operatorname{O\rightarrow Pb}})$ is close to zero.
Similarly, the formation of Pb vacancy in close proximity to pre-existing 
O vacancy is energetically more favorable than single double ionized $V^{\operatorname{Pb}(2-)}$. 
However, the gain in energy is smaller than in the previous case:
$\Delta E^f(V^{\operatorname{Pb}(2-)})-\Delta E^f(V^{\operatorname{Pb\rightarrow O}})$=1.96 eV 
and because $\Delta E^f(V^{\operatorname{Pb\rightarrow O}}) >$ 0,
the lattice distortion induced by O vacancy is not sufficient to make a process of origin of the Pb vacancy spontaneous. 

\begin{figure}
\includegraphics[scale=0.74]{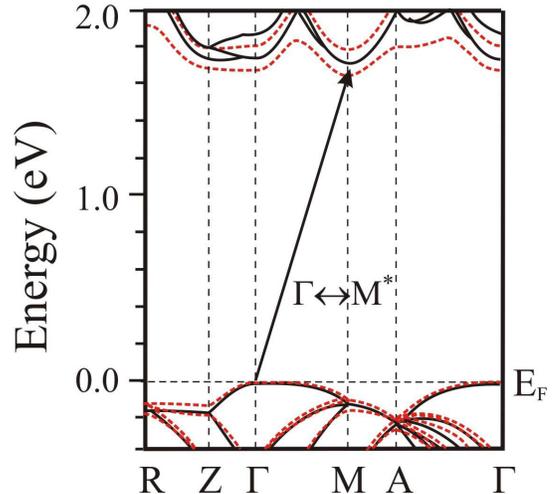}
\caption{\label{fig:fig3} 
Band diagram for ideal supercell (black solid line) and
for the same supercell containing the $V^{\operatorname{Pb-O}}$ defect (red dashed line).}
\end{figure}

To understand an effect of passivation on the transport properties,
the next step is to examine alteration in the electronic properties
induced by the neutral vacancy pair. We found, since the uncompensated
dangling bonds are absent at the $V^{\operatorname{Pb-O}}$ defect site,
the vacancy pair has a little effect on the band formation as shown in Fig.~\ref{fig:fig3}. 
The vacancy pair does not seems to generate any localized state within the band gap, 
but rather induces a slight modification into the band behavior. 
The top valence band and its dispersion are almost unaffected by the appearance of the vacancy pair and, 
therefore the hole microscopic mobility is predicted to remain the same
(it is already low as a result of heavy holes  $m_h^*$=2.44 $m_0$ \cite{masses}).
In the conduction band, the vacancy pair induces a shift of the lowest band down on the energy axis. 
This band is a product of closing of the dangling bonds of vacancies on each other, i.e. its
antibonding orbital. Moreover, it shows a slightly weaker dispersion at the 
$\mathbf{M}^*$ point that is anticipated to cause a reduction in the electron mobility. 

\begin{figure}
\includegraphics[scale=0.37]{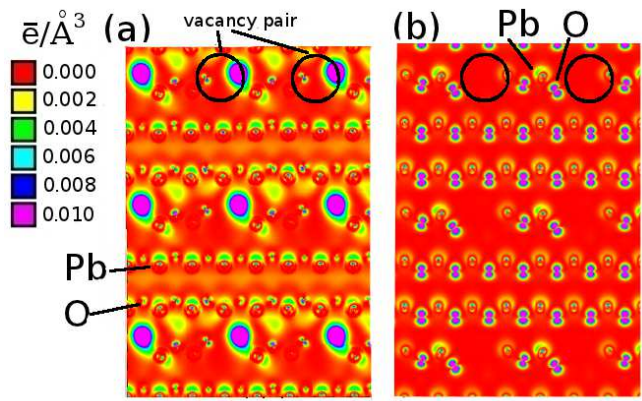}
\caption{\label{fig:fig4} Colour on-line. 
The electron density map (e/A$^{3}$) plotted with 
XcrySDen \cite{xcrysden} for 3$\times$3$\times$3 supercell:
the cutting plane is perpendicular to layers and passes 
through the vacancy site. (a) the conduction band ($E_C$+0.076 eV), (b) the valence band ($E_V$-0.076 eV).}
\end{figure}

To gain insight the band behaviour we have plotted the electron
density map for bottom of the conduction band ($E_C$) and top of the valence band ($E_V$)
in Fig.~\ref{fig:fig4} (a) and (b), respectively. 
The O vacancy was created next to the Pb vacancy. 
The electron density map is plotted for the narrow energy range 
($E_C$+0.076 eV) and ($E_V$-0.076 eV) to catch $E_C$ and $E_V$ in vicinity
of the $\mathbf{M}^*$ and $\boldsymbol{\Gamma}$ points, respectively.
For the conduction band, the electron density shows the
pronounced localization in the layer containing $V^{\operatorname{Pb-O}}$ and, in particular,
on the atoms around the defect site: continuity of the electron network is
disturbed that should affect the electron microscopic mobility.
For the valence band, an alteration in the electron density is
confined entirely to the vacancy site where density vanishes.
Although such hole in the electron density can affect the carrier
transport locally in the vicinity of the defect,
beyond that density remains undisturbed, suggesting that the hole mobility is not altered.

From our study, one can conclude that Pb vacancies in 
PbO layers are passivated by O vacancies to form the neutral vacancy pair. 
It seems reasonable to assume that negative effect of the vacancy
pairs on the transport properties of PbO photoconductive layers would be minimized as the
$V^{\operatorname{Pb-O}}$ pair can not act as trapping center
(dangling bonds are absent and extra electron or hole is delocalized). 
Moreover, as pair formation involves merging of single Pb and O vacancies 
($E_{bind}$ =-1.47 eV) which otherwise are active traps,
it is foreseen to suppress the carrier trapping.
As the concentration of O vacancies is higher than that of Pb vacancies,
the large fraction of $V^{\operatorname{O}(0)}$ is
left uncompensated to be filled with two electrons \cite{defects}.
The experimental evidence of $n$-type conductivity in PbO layers \cite{bigelow} 
supports this observation. Our finding offer a practical way to improve
the transport properties of thermally deposited PbO layers.
The post-growth annealing (in vacuum or oxygen atmosphere)
would initiate migration of the O vacancies towards Pb vacancies
with driving forces defined by the vacancy interaction $E_{int}$=0.55 eV
and facilitates their merging. This is expected to reduce an amount of
ionized centres in PbO thus improving drift mobilities of electrons and holes as well suppressing recombination.

\section{Acknowledgement}
Financial support of Ontario Ministry of Research and Innovation through a
Research Excellence Program Ontario network for advanced medical imaging detectors is highly acknowledged.

\end{document}